Article

# Smart Learning in the 21st Century: Advancing Constructionism Across Three Digital Epochs

Ilya Levin [1], Alexei L. Semenov [2,*] and Mikael Gorsky [3]

1. School of Education, Tel Aviv University, Tel Aviv-Yafo 6997801, Israel; ilia1@tauex.tau.ac.il
2. Association of Teachers of Mathematics, Derby DE1 1FR, UK
3. School of Computer Science, Holon Institute of Technology, Holon 5810201, Israel; mikaelg@hit.ac.il
* Correspondence: alsemenov2021@gmail.com

**Abstract:** This article explores the evolution of constructionism as an educational framework, tracing its relevance and transformation across three pivotal eras: the advent of personal computing, the networked society, and the current era of generative AI. Rooted in Seymour Papert's constructionist philosophy, this study examines how constructionist principles align with the expanding role of digital technology in personal and collective learning. We discuss the transformation of educational environments from hierarchical instructionism to constructionist models that emphasize learner autonomy and interactive, creative engagement. Central to this analysis is the concept of an "expanded personality", wherein digital tools and AI integration fundamentally reshape individual self-perception and social interactions. By integrating constructionism into the paradigm of smart education, we propose it as a foundational approach to personalized and democratized learning. Our findings underscore constructionism's enduring relevance in navigating the complexities of technology-driven education, providing insights for educators and policymakers seeking to harness digital innovations to foster adaptive, student-centered learning experiences.

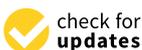

## 1. Introduction

More than half a century ago, at the dawn of the digital age, Seymour Papert pioneered constructionism. Constructionism is a learning theory that emphasizes the construction of knowledge through active creation and experimentation, proposing that learners develop their understanding most effectively when they build personally meaningful artifacts. In Papert's own words: "Constructionism... shares constructivism's connotation of learning as 'building knowledge structures' irrespective of the circumstances of the learning. It then adds the idea that this happens especially felicitously in a context where the learner is consciously engaged in constructing a public entity..." (Harel & Papert, 1991). This innovative educational approach came into life within the context of the emergence of digital technology, as Papert considered computers as a new medium that children could use for making things and expressing themselves (Resnick, 2020). Our paper posits that the ongoing transformation of education in an AI-driven world is inextricably linked to the principles of constructionism. By examining historical foundations and technological advancements, we establish that constructionism offers a valuable framework for effectively integrating AI technologies into educational practices. Moreover, on an intuitive level, constructionism is the most promising educational philosophy for mass schools in the era of using AI in the humanities.

The recent emergence of generative AI brought both exciting possibilities and significant challenges to education, necessitating a re-evaluation of educational theory and practice in our hyperconnected digital age. This paper examines the transformation of education from the early days of personal computing to the present era of AI proliferation.

Since the late 1970s, the digital revolution profoundly influenced society, causing extensive and rapid changes. For context, consider that, in 2000, merely 6.7% of the global population had Internet access; by 2021, this figure surged to over 63% (World Bank, 2023). Throughout this period, we witnessed the rise of technologies that fundamentally reshaped society.

Network technologies played a pivotal role in this transformation, creating a global information system that interconnects individuals and provides easy access to knowledge across various domains. Consequently, people's lives and perceptions of themselves and the world are evolving. Technologies such as Web 2.0, mobile computing, cloud services, and the Internet of Things altered not only our environment, but also our understanding of our place within it and shaped our behavior and thinking.

These changes are particularly significant in the realm of education. The Onlife Initiative, launched in 2013, conducted an in-depth "Reengineering Exercise" focusing on these digital transformations (Floridi, 2015). The interdisciplinary project brought together experts from fields such as philosophy, science, education, technology, history, culture, and sociology. The resulting "Onlife Manifesto" defined the digital revolution as a revolution in human consciousness, with individuals realizing themselves as hyperconnected entities.

Today, in the era of generative AI proliferation, we appear to be experiencing an even more significant shift in collective consciousness (Wellner & Levin, 2023). While the full implications of AI technologies are not yet fully understood, their crucial role in education is undeniable.

Seymour Papert's constructionism offers a valuable lens through which to view these educational transformations. Papert envisioned learning as an active, creative process—a concept that aligns well with the possibilities presented by our digital, AI-enhanced world. As we confront the challenges and opportunities of integrating AI into education, constructionism may provide insights into effectively leveraging these new technologies.

Constructionist principles can enhance learning experiences in today's technology-driven educational landscape. These principles can guide educators and students in adapting to rapidly changing technology and harnessing it to their advantage. By applying constructionism, we can discover innovative ways to utilize AI and other digital means to construct knowledge in creative and meaningful ways while opposing dangers and reducing the risk of technology misuse.

Our paper examines the core principles of Papert's constructionism, delves into the historical foundations of the digital revolution, and explores the positioning of constructionism in relation to the concept of smart education. We will then investigate changes in self-perception and interpersonal interactions in the digital era, examining these shifts across three key periods: the dawn of the digital era, marked by personal computers; the network era, characterized by the proliferation of the Internet and social media; and the present day, dominated by generative AI.

Through this exploration, we aim to demonstrate how constructionist ideas can be applied in contemporary education, leveraging AI and digital tools to create more engaging, effective, and personalized learning experiences. By understanding the historical context and theoretical foundations, educators and policymakers can better navigate the challenges and opportunities presented by AI in education, ultimately fostering a more adaptive and innovative learning environment for future generations.

## 2. The Information History of Humanity and Its Transformations

Information, perception, storage and preservation, and transmission are inherent to living organisms, starting with the most primitive ones. At the same time, it is clear that the main stages in the development of humankind are to a very large extent connected with how a person works with information. In this aspect, information revolutions associated with the emergence of the following three concepts can be distinguished (Semenov & Ziskin, 2023):

- Consciousness;
- Speech;
- Writing.

These revolutions occurred at increasing (one can metaphorically say exponentially increasing) speeds. A hundred years ago, in mathematics and then in technology, a new revolution began—the revolution of artificial intelligence, and the transfer of human intellectual functions to machines.

On a more detailed scale, we can also mention, for example, the Gutenberg revolution—a leap in the means of distributing textual and visual information. We can also talk about the revolutions of photo-cinema, radio, television, and sound recording, which made it possible to transfer the means of perception, storage, and transmission of visual and sound information outside of human beings.

Considering the processes within the revolution of artificial intelligence under such a "historical microscope", we can talk about the revolution of rational AI, when rational functions of human consciousness were transferred to the machine, such as the ability to perform algebraic calculations; the next one can be described as the transfer of intuitive consciousness to the machine, i.e., the revolution of machine learning; and the (so far) last one, which is intensively progressing today, can be termed the transfer of creative consciousness.

We can also highlight the moment when a revolution reached everyone, similar to the revolution of radio and television broadcasting and the Gutenberg revolution.

In The Onlife Manifesto, the period preceding the late 1970s is characterized as the "information prehistory" of humanity. This era marked the time before the advent of the personal computer (PC), which fundamentally transformed information access and management. With the PC, people gained a tangible, unprecedented ability to independently save, copy, process, transmit, and share information in various forms. Consequently, the birth of the PC represents the starting point of humanity's "information history". In just three decades, a substantial portion of the global population became able to engage with information that was not only freely accessible but also integral to everyday existence.

The mid-2010s are identified as the onset of humanity's "information post-history". This period reflects a shift where individuals began to inhabit an interconnected, digital environment, extending beyond the confines of their physical selves and beginning to perceive themselves as informational entities who are integrated within an expansive informational space.

With the release of ChatGPT 3.5 in late 2022 (or an alternative date of your choosing), the notion of "post-history" underwent a profound transformation, as artificial intelligence entered the forefront of the digital revolution.

The "pre-historical" 1970s represent a time when technology was not yet seen as deeply intertwined with human life. The PC emerged as the "first swallow" of what has since become our digital reality. In this era, individuals with mobile devices enjoy constant, boundless access to a diverse range of information. Today, imagining life without digital technology is nearly inconceivable, and for the younger generation, it is entirely unimaginable. It is almost forgotten, and hard to believe, that during those early days, information technology (IT) was in its infancy, and digital devices were generally remote from everyday people, with computing resources primarily being accessible to specialists, who "shielded" them from the public.

Viewing technology as an extension of the human self was a rare perspective. Although most experts recognized the PC's transformative potential, they tended to regard it predominantly as a technological, rather than humanitarian, innovation—a tool of great importance, yet fundamentally separate from the human experience.

The importance of the PC in the information development of humankind was not immediately realized. The consequences of the advent of the PC extend far beyond technology and are discussed in the following sections. However, here, we explain the uniqueness and fundamental features of the PC compared with its predecessors. Before its emergence, digital information processing was limited to computers that were capable of performing two functions: calculations and control. Unlike bulky professional computers, a personal (i.e., belonging to a human) computer could perform a significantly more extensive variety of functions and execute personal functions rather than industrial ones.

Even the first, not yet perfect, PCs allowed untrained users to write clearly, draw, paint, save, send and receive, and organize text, graphics, and audio information of various kinds. They could also print this information using a printer and play numerous arcade games. Thus, the PC replaced several home devices simultaneously, some of which had yet to become commonplace in households.

At the beginning of the 1980s, the outstanding Russian scientist Professor Andrey Ershov proclaimed that "Programming is the second literacy" (Ershov, 1981). In other words, a new entity would require the younger generation to possess universal skills to work with it. By "Programming", Ershov referred to a concept closely aligned with what, around the same time, came to be known as "Computational Thinking" (CT).

The idea of CT was first articulated by Seymour Papert in his 1980 book "Mindstorms" (Semenov, 2017/in press), where he envisioned computers as transformative tools for learning, exploration, and creativity. In Papert's original conception, computational thinking involved applying computational concepts such as algorithms and procedural thinking to solve problems and develop new ways of understanding. His vision, grounded in constructionist learning theory, emphasized programming as a medium for children to actively construct their own knowledge.

Papert's perspective marked the beginning of a new era in which humans could engage intensively and individually with technology, free from intermediaries. However, even this forward-thinking view of computers only reflected the scope of computational thinking as it was understood in the 1980s.

In 2006, Jeannette Wing significantly expanded the concept in her influential paper (Wing, 2006), reframing CT as a fundamental skill for everyone, not just computer scientists. Wing defined CT as an approach to problem solving, system design, and understanding human behavior, drawing on core computer science principles such as abstraction, decomposition, and automation.

Since then, our understanding of computational thinking continued to evolve, shaped by technological advancements. The advent of smartphones, for instance, revolutionized how people interact with technology. Unlike the era of notebooks, smartphones brought oral and written communication, Internet access, and automatic translation, both written and spoken, into the hands of the general population. This represented a quantum leap; previously, most people did not carry tools such as keyboards, tape recorders, distant phones, or text communication devices with them. Now, these capabilities are integral to daily life.

Another interpretation of the changes taking place is the view that people gained access to information resources to the same extent that they previously accessed electric power and some natural resources. This gave rise to the concept of a "knowledge economy" (World Bank, 2007). The fundamental difference from natural resources is that people do not obtain information resources externally but create them themselves. For these resources to become truly publicly accessible, "reusable", and generally significant, the so-called "autoformalization" of professional knowledge becomes necessary in many situations. In other words, providing individuals with the ability to express their professional skills in algorithmic form is a critical competency, and education should develop in this direction (Castells, 1996b). This interpretation of the digital revolution is no longer purely technological, it is socio-economic in nature.

It is important to note that both these approaches treated the PC as a revolutionary phenomenon and proclaimed an era of universal "computer literacy".

In the late 1970s, Seymour Papert offered his interpretation of the new technological reality and the educational prospects opened by the advent of IT, primarily through the birth of the PC (Papert, 1980b). He understood before others that the PC was not just a new computing device, but an entirely new object, capable of transforming humanity's knowledge of the world and attitude toward it. This fresh and unexpected view was the starting point of his ideas in the field of education.

Papert saw the emergence of digital technologies as the beginning of future changes in people's worldviews, an essential element of inner human culture. Based on this understanding, he developed a new concept of education called constructionism (Papert, 1980b).

As a student of Piaget, Papert regarded Lev Vygotsky's contributions as foundational to the educational philosophy of constructionism. In 1931, Vygotsky himself described the history of humankind's revolutions in information technologies and the transformative effects these revolutions had on humanity itself (Vygotsky, 1980). In a sense, Vygotsky anticipated the very revolutions discussed in this section, foreseeing the sweeping shifts that new information technologies would bring to human experience.

## 3. The Digital Revolution as a Transformation of the Human Worldview

More than a hundred years ago, Sigmund Freud described three, in his view, key revolutions in people's worldviews (Freud, 1955).

The first revolution, according to Freud, is the revolution of Copernicus, which abolished humanity's ideas regarding the location of the Earth of Man in the center of the Universe, placing the Earth in the series of planets of the solar system (which continued in subsequent cosmogony and cosmography). The second is Darwin's revolution, which abolished Man's opposition to all other living things, placing humans in the general evolutionary flow (which continued in genetics, etc.). The third is the revolution of Freud, which abolished the idea of the human mind as controlling the behavior of Man (and continued in the ideas of the collective unconscious and mass consciousness).

According to this logic, we live in the era of the next revolution in the history of humankind, when humans cease to view themselves as separate subjects and inhabitants of the physical, natural world, and become inhabitants of the information cyberspace (a hyperconnected information organism). Moreover, in this revolution, the idea of the biological brain of humans as the only substrate of intellectual activity has been abolished. This intellectual activity is today shared by humankind with artificial intelligence, supporting the activity of each of us and communication between us. We can call this the Turing revolution, which shows that humans are not the only carriers of rational thinking. This revolution was preceded by another, which can be called the Lobachevsky–Goedel revolution, which showed that the whole of humanity would never comprehend the "Final truth" about the world and answer every question of "How is it really". The development and widespread use of IT have enormous ethical, legal, and political implications. IT is not just a set of tools; it is a powerful influence on our environment that increasingly affects the most important components of our worldview: our perception of ourselves (who we are); our interactions with others (how we communicate); our concept of reality (our metaphysics); and our interaction with reality (our activity). These transformations were considered in relation to the information technologies of his time and before by Vygotsky (1980).

The transformation of the human worldview in the digital era can be analyzed through Couldry and Hepp's concept of a 'mediated construction of reality' (Couldry & Hepp, 2016). This framework highlights how technologically mediated processes fundamentally shape our perception of reality. In an era of 'deep mediatization', media technologies and infrastructures are not mere supplements to reality, but actively construct how we understand and navigate the world. This mediated construction is not merely an overlay on some pre-existing reality, but rather constitutes the fabric of how we understand and navigate our world, especially in educational contexts where knowledge construction is increasingly dependent on digital tools and environments.

Complementing this perspective, Citton's ecology of attention framework (Citton, 2019) explains the transformation of cognitive engagement with reality. Rather than viewing attention as an individual resource, this ecological approach situates it within broader technological and social conditions that shape perception, learning, and interaction. In educational contexts, students' attention patterns are not solely individual choices, but are influenced by the media environment. Recognizing this influence highlights the need to intentionally design these environments to foster more effective learning. This perspective aligns with the view that digital technologies are not mere tools for enhancing traditional education, but integral to constructing contemporary educational realities.

We assert that constructionist ideas correspond to changes in worldview that characterize the digital revolution and propose a model that aligns the main ideas of the constructionist approach with these components of a worldview.

Based on this model, we will analyze the digital transformation of school education from the 1970s to the present day. Specifically, we discuss two components of the human worldview: our perception of ourselves and our interactions with others.

Below, we will demonstrate how these components of worldview changed under the influence of these transformations, tracing their evolution from the inception of Papert's constructionist approach, through the formation of a network society, to the current stage associated with the emergence and spread of artificial intelligence. Today, it is evident that Seymour Papert was one of the first thinkers to recognize IT as a unique means of cognition designed to transform both ourselves and our perception of the world. As mentioned earlier, this recognition forms the initial premise of this work, which aims to show that the approach developed by Seymour Papert and his associates contains the beginnings of a new reality and our place within it in the digital age.

At the same time, when discussing the evolution of Papert's ideas in the digital age, as well as the evolution of the ideas of his friend and colleague Marvin Minsky (Minsky, 1988), we are not asserting that he foresaw this exact development of information technologies and society. Instead, we focus on preserving and developing these insightful and valuable ideas as the foundation of today's education, as well as their prospects for the future.

## 4. Changing Self-Perception: Personalization and the Expansion of the Personality

In this section, we explore the transformation of individuals, their roles in the world, and their understanding of new opportunities in the digital age. These questions constitute the first and most essential components of a worldview that we explore in our work.

It is understandable that the PC, as a personal tool, would become the foundation of a new media reality. The PC's role in changing the traditional media environment that was established during the printing revolution is difficult to overestimate. Individuals gained the opportunity to transfer their information and communication activities to electronic media, opening up previously inaccessible ways of creating, storing, transmitting, and processing information of various forms and types. This alone represented a significant expansion of personal capabilities.

This expansion of human capabilities was instrumental. It was centered on replacing traditional "printed" methods of information activity with new, electronic ones.

At the dawn of electronic media, Marshall McLuhan, a pioneering figure in media studies and an influential Canadian philosopher, proclaimed in his 1964 book *Understanding Media: The Extensions of Man* (McLuhan, 1964) that "The medium is the message". He suggested a radical departure from traditional views of media as passive tools. Media, according to McLuhan, rather than merely enhancing human abilities or optimizing existing processes, actively shape and transform human perception and experience, creating entirely new capabilities and ways of interacting with the world—capabilities that had no equivalents in previous media eras.

This aspect of non-instrumentality is particularly evident in the digital media of today, which fundamentally redefines the scale and nature of personal and social expansion in the digital era. Digital technologies, beginning with computers and the networked environments that they enable, introduced phenomena that transcend simple communication or information processing. From immersive digital environments to AI-driven interactions, these technologies enable modes of expression, self-perception, and connection that are unique to the digital age, fundamentally altering how individuals relate to both information and each other (Gallese, 2024). We argue that this non-instrumentality is the defining feature of digital media, shaping its revolutionary impact on the modern human experience.

From our perspective, the first non-instrumental phenomenon of the digital era is the emergence of computational thinking as the most important component of cognitive activity for people in general, and schoolchildren in particular. The non-instrumental aspect of computational thinking has no analogs in previous media eras, unlike, for example, a computer keyboard, a text and graphic editor, or even email. Never before has an individual been given the opportunity to formulate the rules for an artifact's functioning without limits, thereby enabling it to operate independently. Moreover, this can be achieved in an understandable language—a programming language. Recognizing the uniqueness and unique role of computational thinking for schoolchildren became the focus of Papert's attention in the mid-1970s and took a central place in his theory of constructionism. Today, in the digital era, computational thinking manifests itself even more vividly. It is not without reason that the leading contemporary researcher of digital media, Lev Manovich (2013), refined McLuhan's famous definition to say that "The software is the message".

It is important to emphasize that when discussing the role of computational thinking in expanding the personality, we interpret the activity of creating software much more broadly than mere programming. It represents a new type of cognitive activity, forming a new human perception of the surrounding world (Papert, 2000). Computational thinking introduces an algorithmic component to our traditional perception of the world, fostering the idea that the phenomena around us are based not only on the laws of nature, but also on algorithms, without which understanding the world is impossible. Therefore, the importance of developing this new type of thinking when educating schoolchildren cannot be overstated, while some researchers view computational thinking via lenses of a broadly applicable competence domain (Yadav et al., 2017).

## 5. Papert's Constructionism

The authors of this paper had the distinct privilege of actively engaging with the pioneering work of Seymour Papert and his colleagues since the late 1980s, integrating his innovative ideas directly into their educational practice and conceptual designs, often under Papert's direct involvement (Semenov, 2017/in press). As is usually the case with visionary figures whose contributions reshape human understanding, S. Papert's ideas were frequently articulated through aphorisms and paradoxes, and his statements could sometimes appear self-contradictory. This tendency toward philosophical fluidity and bold, occasionally contradictory stances reflects Papert's intellectual depth and openness to evolution, allowing his views to adapt and expand over time.

One intriguing example of Papert's complex perspective, which invites agreement only within specific contexts, is his critical stance toward the Internet. Although this position might seem counterintuitive today, it underscores his commitment to fostering active, hands-on learning experiences and his concern that passive information consumption might undermine his championed constructionist principles. As his ideas develop, Papert's occasional skepticism of emerging technologies highlights his nuanced and layered approach to considering the potential impacts of digital tools on education and human development. This openness to reassessing his own views is emblematic of the depth and adaptability that defined Papert's legacy as a thinker and educator.

In this section, we articulate our interpretation of constructionism, grounding it in our concept of the "expanded personality", a notion of the individual that encompasses a symbiotic relationship with external tools and technologies. This expanded personality reflects a human being who extends their cognitive, communicative, and operational capacities through various instrumental aids, from traditional tools such as the clock and pen and paper to the vast resources offered by an encyclopedia or a mobile phone that is connected to the Internet. These tools serve as conveniences and integral components of a person's cognitive framework, fundamentally enhancing the scope and flexibility of thought, memory, and interaction. This perspective situates constructionism within a broader understanding of learning and knowledge building, where the integration of external resources becomes essential to constructing knowledge and expressing personal agency.

Constructionism is an educational philosophy that emphasizes independent creation, building, invention, and discovery of new concepts by students through interaction with others, engaging in collaborative activities, motivation, and scaffolding provided by them. Parallels can be drawn here with social constructivism and co-constructivism, as described by W. Fthenakis (2015a, 2015b).

Building on the insights of John Amos Comenius (1680), Papert underscored the central importance of mathetics—the art, science, and technology of learning—placing it above traditional didactics in guiding students. At the heart of education, he argued, lies the fundamental constructionist goal: cultivating the ability to autonomously build knowledge in its broadest sense. This encompasses not only acquiring knowledge, but also developing skills, competencies, and attitudes that contribute to a well-rounded, adaptable learner.

Central to this constructionist skillset is the capacity to construct and refine one's understanding within the framework of an "expanded personality", in which cognitive tools and external resources are integrated into personal knowledge building. Such a learner is empowered not only to apply their understanding practically, but also to connect it contextually and engage with others effectively. This active, self-directed learning focus exemplifies the shift toward fostering lifelong, self-sustaining intellectual growth.

In constructing and applying knowledge, individuals draw upon a range of material, informational, and digital resources, a framework informed by the theories of Vygotsky and Papert. Knowledge developed within the "biological" component of one's personality—the brain and cognitive faculties—can be transferred to and augmented by the "digital" component, creating a fluid interaction between human cognition and digital aids. This integration allows for the digital aspect to not only store and enhance knowledge, but also to actively participate in its application, thus enriching and expanding the functional reach of personal intelligence. Through this seamless exchange, the digital and biological facets of the individual become interdependent, enabling more dynamic and versatile engagement with knowledge.

The primary source of motivation lies in the pursuit and continuous realization of outcomes that hold value both for oneself and for others—extended personalities interconnected within a larger social and digital ecosystem. Currently, the core of this motivation remains rooted in the will of the biological component of the personality, governed by one's intrinsic goals and aspirations. However, under the guidance of what might be termed a "meta-will", an individual can delegate certain motivational functions to the digital aspect of their personality. This digital component, in harmonious interplay with the biological self, can assist in sustaining and even enhancing motivation by offering reminders, tracking progress, and reinforcing commitments, ultimately supporting a more integrated approach to personal and collective achievement.

The educational process assumes that skills and motivation are acquired during construction. This construction aims to solve important tasks for the student, often arising from interactions with essential others such as friends, parents, or teachers. Key examples of such tasks include challenging problems—issues the learner does not know how to solve. Addressing these problems may be helpful or even necessary in enhancing technical skills. However, such skills, such as high-quality manual technical drawing or pen-and-paper calculations, are becoming less common.

Other tasks involve situations where a student can achieve individually high and relevant technical results: playing a musical instrument, performing a high jump, cycling without hands, memorizing a significant poem, doing mental arithmetic, practicing calligraphy, or acquiring encyclopedic knowledge. When improving in such tasks, challenging problems may arise; often, everything begins with such a task, such as learning how to ride a two-wheeled bicycle. To achieve the primary goal, we can choose contexts and limit the means that are available to the student.

Broadly, learning can be understood through two primary approaches (Papert, 1980a):

(1) Instructionism: A "top-down" approach, where learning begins with the acquisition of explicit knowledge. This structured, formal process provides learners with clear information and principles, which they then apply in practice to develop implicit understanding and proficiency. Here, practical knowledge builds upon a foundation of theoretical insight.

(2) Constructionism: A "bottom-up" approach that emphasizes learning through experience and experimentation. In this framework, learners acquire implicit knowledge first, often through trial and error, and later formalize this understanding as explicit knowledge. This sequence allows learners to build theoretical frameworks that are grounded in direct experience and hands-on engagement.

These approaches to learning—top-down and bottom-up—are both contradictory and complementary.

Formal education and research historically favored top-down learning, providing structured content and explicit knowledge acquisition. This approach is deeply embedded in culturally constructed systems of schooling, apprenticeships, and other instructional methods, making it the predominant model within society. However, bottom-up learning is ontologically and ontogenetically more foundational to human development.

Ontologically, bottom-up learning is essential, as top-down learning can only occur with the prior establishment of foundational, conceptual knowledge gained through experiential, implicit learning. Ontogenetically, this precedence is reflected in human development, where children first acquire sensory motor skills and then, grounded in these skills, gradually build explicit knowledge.

The tension between these two approaches became especially prominent with the rise of the digital era. At this juncture, Seymour Papert introduced the pivotal idea that "instructionism"—the traditional method of providing students with fixed content to memorize—should give way to "constructionism". This alternative emphasizes students' active, self-directed construction of knowledge within a learning environment that is enriched by PCs, allowing for a more dynamic and personalized engagement with educational concepts. Papert saw the Turtle's behavior-based learning activity as an example of the constructionist way of learning that was an alternative to traditional "instructionism".

One of the sources of Papert's constructionism was Jean Piaget's constructivism (Papert, 1999). Both approaches, constructivism and constructionism, see the child as "an architect, erecting the structures of his own intellect" through independent creativity rather than receiving "ready-made" knowledge from the teacher. Piaget's constructivism considered this independent creativity primarily in the context of developing predetermined rules and concepts, which ultimately represent generally accepted knowledge, rules, and laws.

Papert's constructionism adapts Piaget's theory to the onset of the digital age. It reflects Papert's understanding of a new form of education, suited to individuals in the digital society.

The constructionist process of learning, which is a sequence of solving specific unexpected problems, is the main goal. The child learns from personal experience, explores the world, and makes their own discoveries rather than knowing what they are told. In new, unexpected situations, they make decisions themselves rather than copying the experience of making decisions in known situations, as is customary in a traditional school. In this sense, Papert goes further in overcoming instructionism than Piaget (Ackermann, 2001).

The core concept of constructionism is "technology-neutral". However, the fact that the birth of constructionism coincided with the beginning of the digital era is not accidental. The computer learning environment, already in the early 1980s, created the impression of the possibility of unlimited creativity, consisting of the creation of various objects, plots, and scenes. According to Papert, the path to their creation is realized through the creation and debugging of a program for a computer and the transfer of the metaphor of programming and debugging to other spheres of intellectual experience and life. It was precisely this sense of the enormous potential of the computer that was the most important inspiration for Papert, his associates, and many schoolchildren of that time, opening for them a previously unknown path to knowledge.

Even the first lessons for schoolchildren with the language of Logo (and even BASIC) in the mid-1980s were amazing. The very fact that students could write a short and clear program for the Turtle to move and run it several times in a lesson during debugging was surprising and unusual for adults and fascinating for children. Many high school students could not be dragged away from the computer. One could often hear them say "I can't even believe it: now we can do everything!" It is no coincidence that this feeling of unlimited possibilities provided by the computer influenced the formation of the current generation of IT specialists. It vividly characterizes the self-awareness of an expanded personality and the uniqueness of computer thinking itself, which is completely new for schoolchildren. Of course, not everyone had the chance to feel this. This, again, confirms the importance of ensuring computer literacy and computer thinking skills in school. At the first stage of computerization, the most important property of the self-awareness of the extended personality was identified: the personalization of the perception of the surrounding world. Papert emphasized this, speaking about the personal, unique, and even intimate nature of humans' cognitive and creative activity. According to his approach, people learn about the surrounding world by creating their own microworlds. Papert describes the fundamental importance of educational activity, which consists of creating one's own microworlds, in the very first lines of his famous work *Mindstorms* (Papert, 1980b). He recalls the "gears" with which he played as a child in the garage and based on which he built his own microworlds. The very process of creating these environments was a process of active learning and, importantly, learning about oneself. As Papert wrote in *Mindstorms*, "It is possible to design computers so that learning to communicate with them can be a natural process, more like learning French by living in France than like trying to learn it through the unnatural process of American foreign language instruction in classrooms".

Described by Papert as "a machine that returns learning to its natural character", the computer becomes a means of cognition when the student, just as young Seymour with gears in the garage, perceives themselves as a creator, an "architect" of new worlds, and at the same time of their new intellectual structures, which is the most important component of the process of cognition. Papert notes the active, individual, emotional, and very personal nature of educational activities of this type. He emphasizes that this activity changes the child's self-perception and strengthens their cognitive abilities, making them significantly more active than a pre-computer-era schoolchild. This is the first, but very important, component of personality expansion.

## 6. Constructionism in Smart Education

Within the broader smart education framework (Zhu et al., 2016; Pozo, 2017), which encompasses all aspects of technology-enhanced education, smart learning represents learner-centered, knowledge-constructing mathetics (Konstantinov & Semenov, 2022). Other components complementing smart learning include smart environments, which deal with designing both physical and virtual spaces that support learning; smart teaching, which focuses on methods for facilitating learning experiences; and smart assessment, which evaluates learning outcomes in these environments.

The paradigm of smart education represents a significant evolution in our understanding of technology's role in the learning process. It embodies a fundamental reconsideration of how technology and learning intertwine, moving beyond the instrumental use of technology to a deeply integrative approach, where technology becomes an organic part of the educational ecosystem. Within this framework, we propose a novel conceptualization: positioning constructionism, a theory developed by Seymour Papert, as the embodiment of smart learning—a core component of smart education.

This positioning, encapsulated in the equation "Constructionism = Philosophy of Smart Learning", carries several important implications for educational practice and theory. It underscores the importance of active, hands-on learning experiences, where technology serves as a medium for exploration and creation rather than merely a tool for routine operations. Constructionist smart learning prioritizes learner autonomy, emphasizing self-directed learning and knowledge construction through experience and creation rather than passive reception of information. They also share a common emphasis on fostering higher-order thinking skills, creativity, and problem-solving abilities.

This conceptualization rests on several key hypotheses. Constructionism, emphasizing learning through creation and active engagement, can be understood as a concrete manifestation of smart learning within the broader smart education paradigm. Perhaps most crucially, we contend that both constructionism and smart education represent a beyond-instrumental approach to educational technology, integrating it seamlessly into the processes of learning and cognition. Furthermore, we posit that smart education is an encompassing concept that subsumes various facets of the constructionist educational process, including smart teaching and smart assessment, which are the two facets of smart feedback.

This beyond-instrumental perspective marks a significant departure from traditional views of educational technology. It suggests a symbiotic relationship between learners, educators, and technology, where the boundaries between these elements become increasingly blurred. In this context, technology is not merely a means to an end, but an integral part of the learning environment, shaping how knowledge is constructed, shared, and applied. The integration of constructionism into the smart education framework thus offers a robust theoretical foundation for understanding and implementing technology-enhanced learning environments that prioritize active, learner-centered approaches to knowledge construction.

By positioning constructionism as the philosophy of smart learning within the broader smart education paradigm, we provide a clear and focused way to understand the role of technology in modern educational practices. This framework bridges the gap between established educational theories and emerging technological approaches, emphasizing the importance of active, creation-based learning in technology-rich environments. As academic institutions continue to adapt to rapid technological advancements, this conceptualization can serve as a guiding principle for curriculum design, pedagogical approaches, and the development of learning environments that truly harness the potential of technology to transform education.

*6.1. The Personalization of Education in the Networking Age*

During the early era of personal computing, the impact on human cognition was less apparent. At that time, discussions about computational thinking—the defining characteristic of the period—remained largely confined to academic and educational circles. In the network era, the expansion of personality becomes a subject of discussion for the general public, and most importantly, is intensified due to the intensive development of Internet technologies and the formation of a network society based on these in "The Information Age" trilogy (Castells, 1996a, 1997, 1998).

The hyperconnectivity of its inhabitants characterizes the network society. Everything and everyone are connected to each other: people with people, people with machines, and machines with machines. In these conditions, the microworlds of people turn out to be atoms of the hyperconnected world. This circumstance creates a new dimension in the self-perception of the extended personality. In the network, an individual's extended personality is formed and developed not only by creating its own microworld, but also in intensive interaction with other personalities and their microworlds. A person finds themselves in a world of numerous interactions, where their personality is identified and functions. This new type of personality formation occurs in the context of a new phenomenon of personal online identification (personal identity online) (Floridi, 2011; Rodogno, 2011). This latest extension of the personality, associated with hyperconnectivity, provided every inhabitant of our planet with the possibility of unlimited access to other inhabitants (e.g., via social networks), to any place on Earth (e.g., via Google Maps), and to any piece of knowledge (for example, using Wikipedia) (Serres, 2012).

Interaction with others, with children, and with a teacher in the creation process is an essential component of learning, a source of motivation, and a fundamental element of constructionism for Papert. He emphasized that the worlds created by the child and the objects in them are interesting and significant both for the child and for other people. This social aspect of constructionism, highlighted by Vygotsky (and Papert pointed to Vygotsky as a source), was developed by Wasilios Fthenakis in his concept of co-constructionism (co-creation) W. Fthenakis (2015a, 2015b).

The three most important human abilities, being memory, imagination, and reason, have been the subject of well-known philosophical reflections and discussions (Hobbes, 2010; Kant, 1781; Hume, 1882). Each of these human abilities found its place in the networking era.

Our memory has been greatly expanded, and sometimes even replaced, by cloud technologies that provide convenient network access to shared storage of customizable information and computing resources on demand. The feeling of expanding our memory in the network era is becoming the norm in our lives.

Digital technologies enriched human creative expression, amplifying our imagination across all realms of spiritual culture, from visual arts to music. The emergence of 'remix' art exemplifies this transformation: a genre where creators blend existing media fragments with digital effects to produce novel multimedia works. This accessible technique democratized artistic expression, finding enthusiasts across diverse groups, from seasoned artists to young students.

Under these conditions, the self-perception of the extended personality undergoes a change. It shifts towards connections and interactions in which the person takes part. This is one of the digital transformations of society mentioned above; namely, the transition from the primacy of autonomous subjects and objects to the primacy of interactions, processes, and networks of expanded personalities (Ess, 2012).

It is obvious that the self-perception of the network (online) personality and of the habitual (offline) personality differ. For Papert, personalization of the environment was associated with two types of microworlds: physical structures and computer simulations of that which is physical and imaginable. In his works and public lectures, he allowed for and foresaw the emergence of mobile personal environments, such as miniature computers and wearable technologies. However, devices connected to global networks were not, and could not be, discussed in the 1980s. In contrast, it is precisely mobile network technologies that bring the activation of self-perception to a new level. In 1996, Papert published a book, Connected Family: Bridging the Digital Generation Gap (Papert, 1996), the very name of which opposes the idea that digital technologies are divisive.

In today's networked society, a person is not only organically and continuously connected to the network; they are also in their own personal information environment. This personalization of the environment reached a new level, owing to personal mobile devices, which provide an environment with unique properties of contextual and social awareness (Wellner & Levin, 2023). Due to the presence of these properties, in addition to personal identification, the personal environment (based on a mobile device) allows for the consideration of the location of the individual, their life history, their cultural context, and the influence of others. It is important to note that such personalization of the environment occurs automatically, without the participation of the person themself. Thus, the basic level of personalization is formed: passive personalization.

In the digital era of information abundance, in addition to passive personalization, active personalization is also formed. The personal environment, being part of the global in-fosphere, connects a person with an abundance of information resources of the most diverse nature. At the same time, they have the opportunity to choose means that match their taste, desire, and mood. The choice of desired content—instruments, sources, and media—is made by the person themself. Unlimited access to content leads to the individualization of the personal cultural space. People have the opportunity to choose their desired content in the process of study, work, and entertainment. As a result of such active personalization, the personality forms its own unique and changing cultural digital component of their expanded personality, intensively interacting with the expanded personalities of others. Moreover, the expanded personality is formed in accordance with its inclinations, abilities, and priorities. Let us recall that this process is currently supported by networked AI tools.

It can be argued that the expansion of personality in the digital (network) era significantly changes both the (biological) personality itself and its self-perception. Research into this phenomenon and its consequences is of great interest today to anthropologists, sociologists, psychologists, philosophers, and, of course, researchers in the field of education.

*6.2. The Personalization of Education in the Age of Generative AI*

From a constructionist perspective, personalization has been a critical component of education since the dawn of the digital age (Lukowicz et al., 2012). This role has not only been confirmed, but also strengthened in the networked age. However, with the advent of generative AI, personalization is reaching an entirely new level, emerging as an unprecedented phenomenon. It is not merely the interactivity of generative chatbots that is striking, but the way that they create the impression of intentionality in the interlocutor (the student). Interaction with generative AI is no longer just an exchange of information; chatbots become personal interlocutors that are capable of generating content that is deeply tailored to the student's needs. The knowledge produced reflects the individual's unique way of thinking, as well as responding to and shaping their questions. Over time, the chatbot evolves into a personal interlocutor, functioning as a kind of "alter ego" for the student. This interaction takes students beyond traditional methods of analysis and comprehension, immersing them in a personalized, creative process. Furthermore, this innovative personalization extends to the selection and creation of educational content. This means that it is the full-scale implementation of the dialogic learning of Rupert Wegerif (Phillipson & Wegerif, 2017; Wegerif & Major, 2023) and Mikhail Bakhtin's dialogic philosophy (Bakhtin, 1986).

The role of generative AI in personalization can be examined through the lens of cognitive load theory (Sweller, 2011), which is especially relevant in the era of AI-enhanced learning. By adapting its output to match the learner's expertise level, AI optimizes the intrinsic cognitive load, ensuring that students' cognitive resources are effectively managed. Offloading tasks, such as information gathering or basic problem structuring to AI, allows learners to focus their working memory on deeper conceptual understanding and higher-order thinking.

Additionally, AI reduces extraneous cognitive load by presenting information tailored to individual learner characteristics and prior knowledge, while sustaining germane cognitive load to support schema construction and automation. This aligns with cognitive load theory's focus on managing working memory limitations and element interactivity for more effective learning.

Looking back, it seems that the constructionist approach to learning was designed in anticipation of the kind of interaction we now observe between the learner and the learning environment. A clear example of this is the concept of the "object to think with", introduced and developed by Seymour Papert. His famous gears were an early embodiment of this idea. However, when considering his vision of the computer as an "object to think with", a notable limitation was the reliance on coding and traditional algorithmic thinking as intermediaries between the learner and computer. This aspect of the learning experience lacked the seamless, personal, and intimate interaction characteristic of working with physical objects such as gears.

Today, with the advent of generative AI, the concept of the "object to think with" has undergone a profound transformation, realizing and even surpassing its original potential. This evolution carries significant implications for both knowledge formation and the development of self-identity. Interactions with generative AI involve a two-way process of human–machine co-creation, where new ideas emerge, and reflective feedback on the thinking process is provided (Levin & Tsybulsky, 2017). This enables learners to construct knowledge through a personalized, collaborative, and creative process. It now seems natural to view generative AI as an evolution of Papert's "object to think with", which today becomes a "partner to think with".

## 7. Transforming Human Interaction and Democratizing Education in Smart Education

The second component of our worldview, our interactions with others, in the field of education, refers to the perception of the place of the student in the learning process, at school, in relationships with other students and teachers, and in the education system. The approach proposed by Papert is highly anti-authoritarian, democratic, and therefore, revolutionary. In this particular aspect, Papert's approach has become widely known.

*7.1. Democratizing Education at the Beginning of the IT Era*

Papert notes that the use of computers as a means of teaching is in contradiction with the traditional education system, which is built on a hierarchical principle. In his work "Perestroika and Epistemological Politics" (Papert, 1990, 2016), written under the impression of the events in the USSR and his first visits there, working with Russian teachers and children, he presents evidence that new technologies create a situation in which the education system cannot remain the same. The contradiction that Papert sees in epistemology is the contradiction between two concepts of knowledge: hierarchical, centralized, and distanced from the student, on the one hand, and decentralized and personalized on the other. He characterizes this contradiction as a confrontation between traditional forms of teaching, focused on the teacher, and approaches that consider the student to be at the center of the pedagogical process which is oriented towards their intellectual growth. Papert draws an analogy with the "perestroika" that was taking place in the USSR at that time, noting that Gorbachev's attempt to improve the system, to rebuild it, and to adjust it was not successful. The system collapsed, and in his opinion, the same will happen throughout the world with the familiar centralized education system, with its uniform curricula, textbooks, exams, etc. Papert's call for replacing the old, obsolete education system with a new, decentralized one is a call for the democratization of schools.

This kind of democratization is constituted by the social and primarily ideological content of the second component of the constructionist worldview.

*7.2. Democratizing Education in the Networked Society*

In 1990, ideas of the democratization of education were perceived as utopian dreams that could only be realized in the distant future.

The rise of a networked society has given new momentum to democratic ideals, reshaping both social structures and educational paradigms. This transformation was catalyzed by the Internet's emergence and rapid evolution from a simple communication tool into a pervasive digital fabric that fundamentally altered the global human experience.

With the development of the Internet and social networks, the next stage, hyperconnectivity, emerged and strengthened the democratization process. The ability of learners to connect and share content globally, beyond traditional educational frameworks, became a defining feature of this phase. The constructionist paradigm in education took on new forms through online communities where students could learn from each other, creating and sharing knowledge without mediation by formal institutions. Hyperconnectivity pushed constructionism toward more democratic approaches, softening hierarchies and opening the access to knowledge and creative opportunities for broader segments of society.

The network society is a dynamic society. Communication between people in the network requires a prompt response to any network event. Connectivity is becoming the norm, a natural state of a person. Both students and teachers are now inhabitants of the information cyberspace and not our familiar real world. How to interact, learn, and teach in these new conditions is a difficult question, the answer to which is highly topical.

The process of socialization in the digital world has undergone significant changes. The nature of interaction within the network contradicts the usual hierarchical communication model. In a developed digital society, the network takes interactions between people to a higher level. This level of open and accessible communication contradicts not only the hierarchical structure of traditional society, but even more importantly, the principles of civil society that were formed during the Enlightenment. This problem became the focus of researchers several years ago. A research direction called "Digital Enlightenment" emerged, based on the idea of the inevitability of a new era of Enlightenment, the ideas of which will have a similar impact in the digital age to the impact of the 18th-century Enlightenment on the development of humanity (Baird, 2021). Time will tell whether this optimistic hypothesis is true. Still, today, it is clear that we live in an open world that is becoming our everyday life, and a responsible attitude to this new situation is a requirement of the times. However, the "Digital Enlightenment" had a chance to be much more democratic and open than the authoritarian Enlightenment of 18th-century intellectuals.

Information openness is one of the main features of the digital society. It is clearly manifested in the style of network activity itself, in particular, creative activity. Network inhabitants can now be both consumers and producers of content. In other words, they can be readers and writers at the same time. In the society we are used to, people's creative activity was based on a copyright on the content they create. Authors shared only the results of their work with readers. In other words, they shared their successes. In the digital era, authors often share content in the process of its creation, turning it into a collaboration with readers. Such openness does not contradict the formal principles of copyright as much as the concept of creative activity in the era of print culture, i.e., the culture of the book. The transition from traditional forms of creativity to a more open creative space, where people share content at different stages, characterizes the digital society. The tendency to share content leads to transparency in digital culture. Such "collective creativity" obviously cannot be reflected in the context of the school within the digital society, which will require serious reflection.

At the same time, there is reason to assume that the tendency of openness should balance the tendency of individualization of personality discussed in the previous chapter. The study of the interaction and joint dynamics of individuality and transparency is a subject requiring special attention, since this problem is related to almost all components of education: the curriculum, teaching methods, assessment, etc.

Information openness and the phenomena accompanying it found real embodiment both in pedagogical practice and theory.

Analyzing fundamental works in the field of education (Dewey, 1916; Freire, 1970) reveals three basic principles for democratizing education: subjectivity, redundancy, and cooperation.

1. The principle of subjectivity: In a traditional school, the educational program determines the content of education. The more precisely and in detail this program describes the content and the closer to the program and textbook the teacher conveys this content, the better.

In modern education, the content is individually mastered models of activity in accordance with individual goals, as well as an individually formed system of orientation in the world, allowing these methods to be used effectively. At the same time, society, parents, teachers, and other students participate in choosing these goals, and most importantly, in motivating the child to achieve them.

In digital education, the curriculum does not exist without the student; the content of learning becomes subjective. In other words, the students themselves form the curriculum on the spot, having almost unlimited access to the cultural heritage of humanity. Any knowledge acquires meaning only in the subject. It is understood as a superstructure over personal experience, the result of the personal structuring of this experience. The school ceases to be a place for transmitting knowledge to students and turns into a place for exchanging subjective knowledge between students and between them and the teacher. The function of the teacher in the new school is not to transmit the content of the curriculum, but to organize a variety of activities for students in a new educational environment and to motivate students to learn.

2. The principle of redundancy is necessary for the implementation of subjectivity. Students' personal knowledge is developed as a number of educational goals which unfold in the zone of proximal development of the student and motivates the student to achieve them. Both the general system of goals and the system of tasks, the implementation of which moves towards the next goal, are obviously redundant, as is the entire world around the child and its representation in the digital environment. The child's goals can arise from their own experience and interaction with other people; the education system must respond to emerging requests, providing adequate information sources, the opportunity to consult with specialists, and access to social networks and forums. Such an obviously redundant educational environment allows students to accumulate the necessary experience of activity to develop personal knowledge and build a personal, educational trajectory.

3. The principle of cooperation: Traditional class lessons are based on coercion, and the leading motive in a traditional class is achieving a good grade. A smart school is a school of cooperation. It has equal rights for students and teachers, who have now become equal partners. In such an educational process, the concepts of "teacher" and "student" gradually converge.

The radical democratization of education in the digital age would be impossible without the phenomenon of redundancy and free access to knowledge, reflecting "the transition of society from information deficit to information abundance".

*7.3. Democratizing Education in the Age of Generative AI*

The era of generative AI is the most significant stage of democratization within constructionist education. We are witnessing a radical shift in how technologies are perceived—not just as tools for accomplishing tasks, but as equal participants in the educational process (Levin et al.). Generative AI evolved beyond a mere tool to become an intellectual collaborator, transforming the traditional human–machine hierarchy into a partnership of equals. In education, it functions not as a subordinate instrument, but as an active co-creator, engaging with students in meaningful dialogue. This shift marks a new phase in the democratization of knowledge creation and learning.

Generative AI is not just a technology; it is a new "inhabitant" of our planet, a new agent with which students can interact at a level that is comparable to interaction with a person. AI evolved into a sophisticated educational collaborator—embodying Papert's concept of an 'object to think with' (Papert, 1980b)—enabling intellectual partnerships of unprecedented depth. Unlike previous digital technologies, such as computers and the Internet, generative AI changes the very nature of the interaction, which leads to a novel and unprecedented level and nature of democratization.

This transformation is closely linked to the ideas of the influential French philosopher Gilbert Simondon, who advocated rethinking the hierarchical structure of human–technology interactions (Simondon, 1958). He emphasized that the traditional view of technology as a subordinate tool no longer reflects reality. He discussed how society evolved to treat animals respectfully, recognize equality between men and women, and promote racial equality. This represents a rejection of old systems of control and a transition to a new form of respectful interaction, with technology as an active participant in human life. His ideas are particularly relevant in the context of generative AI, where technologies become integral parts of the cognitive process, not merely objects of manipulation.

Bruno Latour's works, especially his actor–network theory (Latour, 2005), encourage a rethinking of the role of technology and objects in human society. Latour argued that technologies cannot simply be instruments controlled by humans; they are equal actors within a network of interactions. This idea supports the concept of democratization in education, where technologies such as generative AI play active roles in creating knowledge and facilitating teaching.

Another dimension for development may be the digital AI component of the teacher's personality. This component helps each student co-create and dialog with the teacher. They also try to consider the parents' position to form an individual system of goals, as well as tasks that lead to these goals. A system of goals that is possible for different students is included in the smart environment and corresponds to the redundancy principle. However, the concept of "excess" is losing its meaning today.

The digital component can control in detail all the parameters of the students: attention and fatigue in the classroom during the lesson or the deviation of the student's results from those previously planned by the student themself on the way to achieving an individual goal. Thus, AI participates in the implementation of the principle of subjectivity.

This component can draw the attention of the biological part of the teacher to the student's problem, as well as provide a visualization of any degree of detail or enlargement of the situation in the classroom. It can predict the student's results based on their individual history of learning and highlight the most significant elements of this history—the "starry clock of learning".

This is not limited to the recording and predictive role of AI. AI can also take over feedback functions, such as a typical teacher's reaction to a student's actions and results.

The role of artificial intelligence in promoting cooperative principles is evident: it facilitates communication among students in digital environments, records their interactions through audio and video, and analyzes this communication during collaborative work sessions. Thus, the democratization of education in the twenty-first century reaches a qualitatively new level. Within the framework of constructionism, it manifests in the development of intellectual democracy, where both students and technologies collaboratively create knowledge and content. This collaboration weakens hierarchical relationships and fosters more equitable forms of interaction. Consequently, education ceases to be strictly hierarchical, and technologies become active participants in learning and knowledge formation.

## 8. Personalization and Democratization in Generative AI Enhanced Classroom

In the previous sections, we proposed two key directions for integrating GenAI into education: personalization and democratization. In this chapter, we discuss the potential avenues for practically implementing these strategies, as presented in the article, within the school learning process.

A promising approach to personalization lies in creating a learning environment based on the "partner to think with" concept highlighted in the paper. We believe such a kind of environment can be realized through a flipped classroom model, in which students interact with a personalized assistant at home and then discuss their outcomes with teachers in class. However, relying solely on the flipped classroom model is not sufficient to achieve a fully personalized environment. A crucial additional component may involve use of "GPTs" (custom versions of ChatGPT) (OpenAI, 2023), which will enable teachers to develop individualized learning ecosystems tailored to students' needs. This innovation promises a flexible and universal learning environment with significant potential for further research and development. By combining the flipped classroom model with GPTs, educators can actively shape the learning environment through prompt engineering. This integration will produce a versatile setting for student development and open new avenues for investigating and refining pedagogical methodologies.

The article's notion of the intellectual democratization of learning, where generative AI becomes a partner to students in the classroom, can be actualized through multimedia learning materials. This strategy enables learners to express their ideas across various media formats, making the educational process more flexible and inclusive.

Multimedia functions not as a superficial attribute, but as a cognitive tool that transforms thinking. By examining how their ideas are adapted across multiple formats (oral text, written form, dialogue, and visualization), students not only gain a deeper understanding of the material, but also become aware of how differing modes of expression influence the perception of meaning. This reflective process, situated at the intersection of media and content, constitutes a learning act. It equips students to discern the structure of their own thoughts, the contexts in which those thoughts are received, and how best to adapt them for different audiences. For example, when an idea is translated into a dialogue format with the assistance of generative AI, it can become more accessible without being reduced to superficiality. In this sense, we observe a "democratization of ideas" through mediated forms of representation.

Incorporating visual elements or adopting multimodal approaches (e.g., combining text, audio, and images) can amplify this multidimensional perception. In such an environment, AI once again serves as a "partner to think with", stimulating learning by prompting the learner to consciously engage with the transformation of meaning.

## 9. Conclusions

This study dives into the profound coherence between technology and education, viewed through the visionary lens of Seymour Papert's constructionist theory. We unveil the dramatic metamorphosis of educational landscapes by charting three pivotal epochs: the birth of personal computing, the Internet revolution, and the recent explosion of generative AI.

In 2021 we have published the "Charter for the Digital Way of School" (Strauss & Levin, 2021). The pedagogical principles outlined in the charter remain enduringly relevant, providing a stable foundation for educational philosophy in the digital age. However, the charter's technological elements are expected to evolve continually, introducing new opportunities and challenges that will shape and refine our vision of education.

Drawing on the framework of the "Onlife Manifesto", our analysis centers on human perceptions of technology, focusing on two essential dimensions: self-perception and the navigation of complex networks of human interactions in the digital landscape.

Our findings confirm the foresight embedded in constructionism, highlighting its enduring relevance as a vital tool for understanding the interplay between technology and education. This work offers fresh perspectives for educators, policymakers, and technologists and urges a reevaluation of education's role and form in the digital era. The "Charter of the Digital Way of School" provides an educational framework based on these.

Furthermore, our analysis highlights the potential for integrating generative AI technologies and identifies the most promising directions for innovation in educational practices—directions that will undoubtedly require further research and development.

The transformative potential of generative AI in education lies not only in its ability to enhance personalized learning environments, but also in its capacity to redefine the relationships among learners, educators, and knowledge itself. From a constructionist perspective, generative AI must be understood as an active "partner to think with", reshaping the educational paradigm in profound ways.

To realize this potential, the integration of generative AI must move beyond technological adoption to systemic reform. GPTs can serve as a means for educators to craft personalized learning experiences that respond dynamically to students' needs. Similarly, the flipped classroom model (Milman, 2014) illustrates how generative AI can support learners' engagement with content outside traditional settings, fostering deeper exploration and collaborative inquiry within the classroom. These examples underscore how the use of generative AI aligns with constructionist ideals, emphasizing active, meaningful, and contextually situated learning.

However, the inclusion of generative AI in education also demands a parallel reform in digital literacy. This literacy must encompass a nuanced understanding of generative AI's mechanisms, limitations, and societal implications, forming a cornerstone of broader digital competencies. Developing AI literacy as part of students' digital literacy is imperative for their ability to engage critically and constructively with this technology. Without this foundation, the integration of generative AI as Papert's "object to think with" becomes superficial, failing to fully harness its potential for cognitive and creative empowerment.

Moreover, the rise of generative AI prompts an epistemological shift in education. Traditional modes of knowledge acquisition give way to dynamic processes of knowledge creation, necessitating new pedagogical strategies and teacher training paradigms. Educators must be prepared not only to integrate AI effectively into curricula, but also to guide students in navigating the ethical, cognitive, and collaborative dimensions of learning with AI.

Finally, the role of multimedia in education takes on renewed significance in the generative AI era. The ability of AI to generate rich, multimodal content (Digital Transformation, 2024) invites a rethinking of how information is presented and understood, making classrooms more interactive and inclusive. At the same time, the interaction between students and AI in these spaces introduces a new layer of relational dynamics, requiring critical reflection on the implications for learning and social development.

The inclusion of generative AI in education calls for a constructionist reformulation of the educational system, where students and teachers co-construct knowledge through meaningful interaction with AI tools. This endeavor requires not only technological adaptation, but also a profound rethinking of educational values, practices, and literacies. Future research must continue to explore these dimensions, ensuring that generative AI fulfills its promise as an empowering force for education.

**Author Contributions:** I.L., A.L.S. and M.G. have contributed equally to this editorial. All authors have read and agreed to the published version of the manuscript.

**Funding:** This research was funded by National Academy of Sciences of the Republic of Kazakhstan, grant number AP19680007.

**Institutional Review Board Statement:** Not applicable.

**Informed Consent Statement:** Not applicable.

**Data Availability Statement:** The raw data supporting the conclusions of this article will be made available by the authors on request through correspondence author.

**Conflicts of Interest:** The authors declare no conflict of interest.

# References

Ackermann, E. (2001). Papert's constructionism: What's the difference. *Future of Learning Group Publication*, *5*(3), 438.

Baird, I. (2021). Introduction: "Speaking to the eyes"—Reassessing the enlightenment in the digital age. In *Data visualization in enlightenment literature and culture* (pp. 1–27). Palgrave Macmillan. [CrossRef]

Bakhtin, M. M. (1986). *Speech genres and other late essays* (Slavic Series: No. 8). University of Texas Press, ISBN 0-292-72046-7.

Castells, M. (1996a). *The rise of the network society, the information age: Economy, society and culture* (Vol. I). Blackwell, ISBN 978-0-631-22140-1.

Castells, M. (1996b). The space of flows. In *The rise of the network society* (Vol. 1, pp. 376–482). Blackwell.

Castells, M. (1997). *The power of identity, the information age: Economy, society and culture* (Vol. II). Blackwell, ISBN 978-1-4051-0713-6.

Castells, M. (1998). *End of millennium, the information age: Economy, society and culture* (Vol. III). Blackwell, ISBN 978-0-631-22139-5.

Citton, Y. (2019). Attentional agency is environmental agency. In W. Doyle, & C. Roda (Eds.), *Communication in the age of attention scarcity* (pp. 21–32). Palgrave Macmillan.

Comenius, J. A. (1680). *Spicilegium didacticum: Artium discendi ac docendi summam brevibus praeceptis exhibens*. Typis Christophori Cunradi. Available online: https://books.google.ru/books?id=XwP2uhkAvHsC&source=gbs_book_other_versions (accessed on 29 October 2024).

Couldry, N., & Hepp, A. (2016). *The mediated construction of reality: Society, culture, mediatization*. Polity Press.

Dewey, J. (1916). *Democracy and education*. Macmillan, ISBN 1548554499.

Digital Transformation. (2024). *Exploring multimodal AI and its pivotal role in shaping the future of AI*. Available online: https://www.imd.org/blog/digital-transformation/multimodal-ai/ (accessed on 29 October 2024).

Ershov, A. P. (1981, July 27–31). *Programming, the second literacy* [Paper presentation]. 3rd World Conference on Computer Education (pp. 1–17), Lausanne, Switzerland.

Ess, C. (2012). At the intersections between internet studies and philosophy: "Who am I online?". *Philosophy & Technology*, *25*(3), 275–284. [CrossRef]

Floridi, L. (2011). *The philosophy of information*. Oxford University Press.

Floridi, L. (2015). *The onlife manifesto: Being human in a hyperconnected era*. Springer Open, ISBN 978-3-319-04092-9. [CrossRef]

Freire, P. (1970). *Pedagogy of the oppressed*. Bloomsbury Publishing.

Freud, S. (1955). A difficulty in the path of psychoanalysis. In *The standard edition of the complete psychological works of Sigmund Freud, (1917–1919): "An infantile neurosis" and other works* (Vol. XVII, pp. 135–144). The Hogarth Press and The Institute of Psychoanalysis.

Fthenakis, W. (2015a). Co-contruction: Methodological and didactic approach without passive participants. *Sovremennoje doshkolnoje obrazovanije = Modern Preschool Education*, *2*(54), 58–65. Available online: https://cyberleninka.ru/article/n/so-konstruirovanie-metodiko-didakticheskiy-podhod-bez-passivnyh-uchastnikov (accessed on 29 October 2024). (In Russian).

Fthenakis, W. (2015b). *Ko-Konstruktion gemäß Prof. Dr. mult*. Available online: https://www.youtube.com/watch?v=R2XzZozipiA (accessed on 29 October 2024).

Gallese, V. (2024). Digital visions: The experience of self and others in the age of the digital revolution. *International Review of Psychiatry*, *36*(6), 656–666. [CrossRef]

Harel, I., & Papert, S. (Eds.). (1991). *Constructionism*. Ablex Publishing.

Hobbes, T. (2010). *Leviathan: Or the matter, forme, and power of a commonwealth ecclesiasticall and civill, 1651* (I. Shapiro, Ed.). Yale University Press.

Hume, D. (1882). *A treatise of human nature: Being an attempt to introduce the experimental method of reasoning into moral subjects & dialogues concerning natural religion* (T. H. Green, & T. H. Grose, Eds.; Vols. 1&2). Longmans, Green & Co. Available online: https://archive.org/details/atreatiseonhuma00grosgoog (accessed on 29 October 2024).

Kant, E. (1781). *Critique of pure reason*. Wikisource. Available online: https://en.wikipedia.org/wiki/Critique_of_Pure_Reason (accessed on 29 October 2024).

Konstantinov, N. N., & Semenov, A. L. (2022). Productive education in mathematical schools. *Doklady Mathematics*, *106*(Suppl. 2), S270–S287. [CrossRef]

Latour, B. (2005). *Reassembling the social: An introduction to the actor-network theory*. Oxford University Press, ISBN 9780199256044.

Levin, I., & Tsybulsky, D. (2017). The constructionist learning approach in the digital age. *Creative Education*, *8*(15), 2463–2475. [CrossRef]

Levin, I., Minyar-Beloruchev, K., & Marom, M. (in press). Generative AI as a cultural phenomenon in the digital transformation of education. In *Applied innovations in information and communication technology*. Springer.

Lukowicz, P., Pentland, S., & Ferscha, A. (2012). From context awareness to socially aware computing. *IEEE Pervasive Computing*, *11*(1), 32–41. [CrossRef]

Manovich, L. (2013). *Software takes command*. Bloomsbury Academic.

McLuhan, L. H. (1964). *Understanding media: The extensions of man* (1st ed.). McGraw Hill. Reiss; Gingko Press, ISBN 978-1-58423-073-1.

Milman, N. B. (2014). The flipped classroom strategy: What It Is and How Can it Best be Used? *Distance Learning*, *11*(4), 9–11.

Minsky, M. (1988). *Society of mind*. Simon and Schuster.

OpenAI. (2023). *Introducing GPTs*. Available online: https://openai.com/index/introducing-gpts/ (accessed on 29 October 2024).

Papert, S. (1980a). *Constructionism vs. instructionism. Speech to an audience of educators in Japan*. Available online: http://www.papert.org/articles/const_inst/const_inst1.html (accessed on 29 October 2024).

Papert, S. (1980b). *Mindstorms: Children, computers, and powerful ideas*. Basic Books, Inc, ISBN 978-0-465-04627-0.

Papert, S. (1990). *Perestroika and epistemological politics*. Epistemology and Learning Group, MIT Media Laboratory.

Papert, S. (1996). *Connected family: Bridging the digital generation gap*. Longstreet Press. Available online: https://archive.org/details/connectedfamilyb00pape (accessed on 29 October 2024).

Papert, S. (1999). Papert on piaget [Special issue]. *Time Magazine*, *153*(12), 105.

Papert, S. (2000). What's the big idea? Toward a pedagogy of idea power. *IBM Systems Journal*, *39*(3.4), 720–729. [CrossRef]

Papert, S. (2016). The perestroika of epistemological politics. Closing address to the 1990 world computer education conference. *Australian Educational Computing*, *31*(1), 1–11. Available online: https://journal.acce.edu.au/index.php/AEC/article/view/108 (accessed on 29 October 2024).

Phillipson, N., & Wegerif, R. (2017). *Dialogic education: Mastering core concepts through thinking together*. Routledge. [CrossRef]

Pozo, C. L. P. (2017). Smart education and Smart e-learning. In *Conference proceedings of "eLearning and Software for Education" (eLSE)* (Vol. 13, pp. 89–95). Carol I National Defense University Publishing House.

Resnick, M. (2020). *The seeds that Seymour sowed* (Foreword to the anniversary publication of S. Papert's "Mindstorms"). Hachette.

Rodogno, D. (2011). *Against massacre: Humanitarian interventions in the Ottoman Empire 1815–1914*. Princeton University Press.

Semenov, A. (in press). Seymour Papert and us: Constructionism as the educational philosophy of the 21st century. *Voprosy Obrazovaniya/Educational Studies Moscow*, *1*, 269–294, (Original work published 2017). [CrossRef]

Semenov, A. L., & Ziskin, K. E. (2023). Expanded personality as the main entity and subject of philosophical analysis: Implications for education. *Doklady Mathematics*, *108*(4), 331–341. [CrossRef]

Serres, M. (2012). *Petite poucette* (Vol. 125). Le Pommier. 82p.

Simondon, G. (1958). *Du mode d'existence des objets techniques*. Editions Montaigne.

Strauss, S., & Levin, I. (2021). *Charter for the digital way of school*. Available online: https://ort.org/en/charter-for-the-digital-path-of-school/ (accessed on 29 October 2024).

Sweller, J. (2011). *Cognitive load theory: Vol. 55 psychology of learning and motivation* (pp. 37–76). Academic Press.

Vygotsky, L. S. (1980). *Mind in society: The development of higher psychological processes*. Harvard University Press.

Wegerif, R., & Major, L. (2023). *The theory of educational technology: Towards a dialogic foundation for design* (1st ed.). Routledge. [CrossRef]

Wellner, G., & Levin, I. (2023). Ihde meets Papert: Combining postphenomenology and constructionism for a future agenda of philosophy of education in the era of digital technologies. *Learning, Media and Technology*, *49*(4), 656–669. [CrossRef]

Wing, J. (2006). *Computational thinking* (pp. 34–36). Communications of the ACM.

World Bank. (2007). *Building knowledge economies: Advanced strategies for development* (pp. 4–12). World Bank Publications.

World Bank. (2023). *Share of the population using the Internet*. International Telecommunication Union. Available online: https://ourworldindata.org/grapher/share-of-individuals-using-the-internet (accessed on 29 October 2024).

Yadav, A., Good, J., Voogt, J., & Fisser, P. (2017). Computational thinking as an emerging competence domain. In *Competence-based vocational and professional education, technical and vocational education and training: Issues, concerns and prospects*. Springer.

Zhu, Z. T., Yu, M. H., & Riezebos, P. (2016). A research framework of smart education. *Smart Learning Environments*, *3*, 1–17. [CrossRef]